\documentclass[letterpaper, 10 pt, conference]{ieeeconf}  % Comment this line out if you need a4paper

\IEEEoverridecommandlockouts                              % This command is only needed if
\usepackage{graphics}       % for pdf, bitmapped graphics files
\usepackage{epsfig}         % for postscript graphics files
\usepackage{times}          % assumes new font selection scheme installed
\usepackage{amsmath}        % assumes amsmath package installed
\usepackage{amssymb}        % assumes amsmath package installed
\usepackage{color}
\usepackage{array}
\usepackage{accents}

\usepackage{stackengine}
\usepackage{url}
\usepackage{enumerate}

\title{\LARGE \bf Adaptive time delay based control of non-collocated oscillatory systems}

\author{Michael Ruderman% <-this % stops a space
\thanks{M. Ruderman is with University of Agder, 4879 Grimstad, Norway}%
\thanks{This research was supported by RCN grant number 340782.}
\thanks{\textcolor[rgb]{0.00,0.00,1.00}{Author's accepted manuscript, IEEE MED2024}}
}

\begin{document}

\maketitle
\thispagestyle{empty}
\pagestyle{empty}

%%%%%%%%%%%%%%%%%%%%%%%%%%%%%%%%%%%%%%%%%%%%%%%%%%%%%%%%%%%%%%%%%%%%%%%%%%%%%%%%
\begin{abstract}
Time delay based control, recently proposed for non-collocated
fourth-order systems, has several advantages over an
observer-based state-feedback compensation of the low-damped
oscillations in output. In this paper, we discuss a practical
infeasibility of such observer-based approach and bring forward
application of the time delay based controller, which is simple in
both the structure and design. Moreover, robust estimation of the
output oscillation frequency is used and extended by a bias
canceling. The latter is required for positioning the oscillatory
passive loads. This way, an adaptive version of time delay-based
control is realized that does not require prior knowledge of the
mass and stiffness parameters. The results are demonstrated on the
oscillatory experimental setup with constraints in the operation
range and control value.
\end{abstract}

\bstctlcite{references:BSTcontrol}

\newtheorem{theorem}{Theorem}
\newtheorem{rem}{Remark}

%%%%%%%%%%%%%%%%%%%%%%%%%%%%%%%%%%%%%%%%%%%%%%%%%%%%%%%%%%%%%%%%%%%%%%%%%%%%%%%%
\section{Introduction}
\label{sec:1}

Multi-mass systems with an oscillatory passive load are common in
various control applications. A broad class of such systems can be
well approximated by fourth-order dynamics where the first active
body includes the whole actuator plant, while the second passive
body represents the whole oscillating payload to be controlled.
Elastic links, especially those with low damping, make such
systems challenging for control. Moreover, if only the load output
state is available from the sensor measurements, such systems
become non-collocated -- that is the objective of our present
study. For instance in cranes (see e.g. \cite{vaughan2010}) and
winch systems, the vertical vibration dynamics can become
significant due to the elasticities in ropes and cables.
Longitudinal oscillations in hoisting systems (see e.g.
\cite{wang2023}) are also known to be complex, so that the output
oscillation frequency becomes uncertain and valid only close to an
operation point. Likewise, the drill-string systems represent an
exemplary case of oscillating passive loads (with angular motion),
see e.g. \cite{besselink2015}, while such vibration dynamics
becomes even more non-trivial.

In this work, we consider a class of fourth-order non-collocated
oscillation systems (Section \ref{sec:2}), for which stabilization
only a noisy sensing of the load output displacement is available.
We first discuss in detail a practical infeasibility of the
classical observer-based state-feedback design with loop reshaping
by location of the poles (Section \ref{sec:3}). We demonstrate
that even a low measurement noise, in combination with constraints
of the actuator's force and displacement, makes such a
theoretically sound stabilization less usable. Then, it is shown
that the recently proposed time delay based control
\cite{ruderman2021,ruderman23} constitutes a suitable robust
alternative for such class of the systems (Section \ref{sec:4}).
Also, an online adaptation of the oscillation frequency, required
for the control parametrization, is provided, thus extending the
robust estimator \cite{ruderman23}. Experimental evaluation of the
proposed control is shown for the adaptive frequency tuning and
also additional external disturbances (Section \ref{sec:5}).
Conclusions are in Section \ref{sec:6}. Our design of the time
delay based stabilization is simple and relies on consideration in
frequency domain. Here we recall that analysis of systems with
time delay(s) is manageable also with use of the corresponding
transfer functions, see \cite{kao2004}. For analysis of time-delay
systems by means of the signal-norms see also
\cite{fridman2006input}. For tutorial and basics of time-delay
systems we further refer to
\cite{gu2003,michiels2014,fridman2014}.

%%%%%%%%%%%%%%%%%%%%%%%%%%%%%%%%%%%%%%%%%%%%%%%%%%%%%%%%%%%%%%%%%%%%%%%%%%%%%%%%
\section{Non-collocated fourth-order system} \label{sec:2}

\subsection{General framework} \label{sec:2:sub:1}

\begin{figure}[!h]
\centering
\includegraphics[width=0.6\columnwidth]{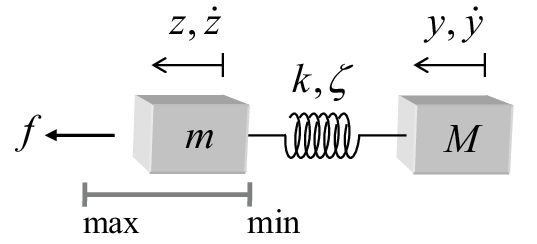}
\caption{General framework of non-collocated system.}
\label{fig:1}
\end{figure}

We consider a general framework of the non-collocated fourth-order
systems as depicted schematically in Fig. \ref{fig:1}. The system
has an active and a passive body with the mass $m$ and $M$,
correspondingly. The relative motion is actuated by the
constrained force $f \in [f_{\min}, f_{\max}]$ and has one degree
of freedom in the shifted (against each other) coordinates
$(z,\dot{z})$ and $(y,\dot{y})$.
Both inertial bodies are connected by an elastic link (i.e.
spring) with the linear stiffness and damping coefficients $k$ and
$\zeta$, respectively. In addition, the first active body can be
subject to the linear damping $\sigma$ (equivalent to the viscous
friction of an actuator), while the second passive body can be
low-damped or even undamped, -- a challenging case that we study
(also with experiments) in this work. Both moving bodies can be
additionally affected by the known constant (or slowly varying)
disturbances $\varphi$ and $\Phi$, each one individually. Further
essential assumption is that the active mass has a constrained
motion range
\begin{equation}
z \in (z_{\min}, z_{\max}), \label{eq:1}
\end{equation}
the fact which is often occurring due to the mechanical limiters
of an actuator. With the above assumptions in mind, the motion
dynamics inside of the range \eqref{eq:1} is given by
\begin{eqnarray}
\label{eq:2}
  m \ddot{z} + (\sigma+\zeta)\, \dot{z} + k(z-y) - \zeta \dot{y} + \varphi &=& f, \\
  M \ddot{y} - \zeta (\dot{z} - \dot{y}) - k(z-y) +  \Phi &=& 0.
\label{eq:3}
\end{eqnarray}
The link damping $\zeta > 0$ is assumed to be relatively low, so
that the load motion of an uncontrolled system experiences a
long-term oscillatory and undesired for applications behavior in
the $(y,\dot{y})$ coordinates, cf. Fig. \ref{fig:3} below. One of
the most challenging characteristics of the system class
\eqref{eq:2}, \eqref{eq:3} is the non-collocation of the control
value $f$ and the single available output of interest $y$. Note
that the measured $y(t)$ state can additionally be corrupted by
the sensor noise. Furthermore, we note that all trajectories in
the four-dimensional state-space of \eqref{eq:2}, \eqref{eq:3} are
continuous and smooth within \eqref{eq:1}, while for $z = z_{\min}
\vee z_{\max}$ the hard switchings appear, see e.g.
\cite{liberzon2003} for fundamentals of the switched dynamics.
This is especially relevant for a constrained motion, addressed
later in sections \ref{sec:2:sub:2}, \ref{sec:3:sub:2}, as a
factor which is inherently limiting the control performance and
feasibility.

\subsection{Experimental case study} \label{sec:2:sub:2}

\begin{figure}[!h]
\centering
\includegraphics[width=0.5\columnwidth]{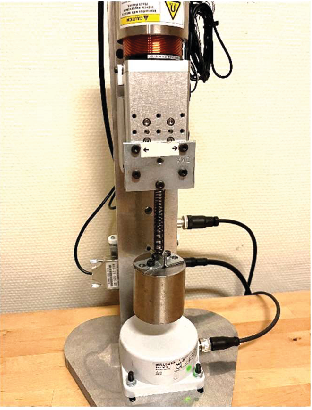}
\caption{Experimental setup of the non-collocated system.}
\label{fig:2}
\end{figure}

\begin{figure}[!h]
\centering
\includegraphics[width=0.98\columnwidth]{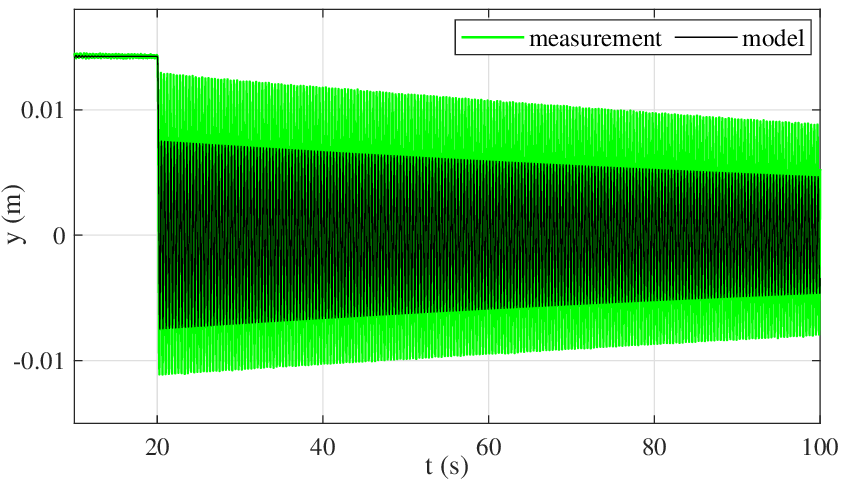}
\caption{Comparison of the measured and computed oscillatory
response of the free fall scenario ($f(t)=0$ for $t>20$ s).}
\label{fig:3}
\end{figure}

The experimental case study under consideration is a
non-collocated 4th-order mechanical system with gravity, which has
a contactless output sensing (Fig. \ref{fig:2}). The actuated body
is the voice-coil-motor with the bounded input and output
$$
u \in [0,\, 10] \hbox{ V,} \quad \hbox{ and } \quad z \in [0,\,
0.021] \hbox{ m,}
$$
respectively. An additional actuator's time constant yields
\begin{equation}
f(s) = F(s)u(s) = \frac{\kappa}{\tau s + 1}\, u(s) =
\frac{3.2811}{0.0012 s + 1}\, u(s), \label{eq:4}
\end{equation}
written in Laplace domain with the complex variable $s$. The
relative displacement of the passive load is measured by an
inductive distance sensor, which has $\pm 12$ $\mu m$ nominal
repeatability and a relatively large level of noise. The latter is
due to a contactless measurement and dynamic misalignments of the
moving body with respect to the inductive field-cone of the
sensor. Here we recall that there is no bearing for the load mass
which is, this way, constituting a free-hanging body, see Fig.
\ref{fig:2}. Further details about the experimental system can be
looked in \cite{ruderman2022,ruderman23}.
For the vector of the state variables $x \equiv (x_1, x_2,
x_3,x_4)^\top = (\dot{z}, z, \dot{y}, y )^\top$, the state-space
model corresponding to \eqref{eq:2}, \eqref{eq:3} is given by
\begin{eqnarray}
\label{eq:5}
  \dot{x} &=& A \, x + B \,f + D, \\
\nonumber y &=& C \, x,
\end{eqnarray}
with the matrices and vectors of system parameters
\begin{eqnarray}
\nonumber  A &=& \left(%
\begin{array}{cccc}
-333.35 &  -333.33  &   0.015     & 333.33  \\
1         & 0     &     0         & 0         \\
0.012     & 266.66 &    -0.012    & -266.66 \\
0         & 0          &  1       & 0
\end{array}%
\right), \\
\nonumber B &=& (1.667, 0, 0, 0)^\top, \quad  C \; = \; (0, 0, 0, 1), \; \hbox { and} \\
\nonumber D &=& (-9.806, 0, -9.806, 0)^\top.
\end{eqnarray}
Worth noting is that the disturbance vector $D$, cf. \eqref{eq:2},
\eqref{eq:3}, is composed by the constant gravity terms acting on
both moving bodies. Further, we emphasize that the system
\eqref{eq:5} has one conjugate-complex pole-pair with the natural
frequency $\omega_0 = 16.4$ rad/sec and extremely low damping
ratio $\delta = 0.031$. The numerical parameter values of the
system model are identified, partially from the available
technical data-sheets of components and partially through a series
of the dedicated experiments, cf. \cite{ruderman2021,voss2022}. An
exemplary comparison between the measured and modeled output
response is shown in Fig. \ref{fig:3} for a free fall scenario.
Starting from non-zeros initial conditions and having $u(t) =
\mathrm{const}$ for the gravity compensation, the control input is
then switched off at $t=20$ sec. This, in a consequence of
$u(t)=0$ for $t > 20$, leads to a fall down of both masses, while
$|\dot{z}| < |\dot{y}|$ due to the actuator bearing. Hence, the
oscillatory behavior becomes largely excited once $z=z_{\min}$.
Note that while the oscillation frequency $\omega = \omega_0
\sqrt{1-\delta^2}$, the damping ratio $\delta$, and the
steady-sate values of $y(t)$ are well in accord between the
measurement and model, the oscillations amplitude is sensitive to
both, the initial conditions and exact knowledge of the moving
mass and stiffness coefficient, cf. Fig. \ref{fig:3}.

%%%%%%%%%%%%%%%%%%%%%%%%%%%%%%%%%%%%%%%%%%%%%%%%%%%%%%%%%%%%%%%%%%%%%%%%%%%%%%%%
\section{Observer-based state-feedback} \label{sec:3}

\subsection{Theoretical framework} \label{sec:3:sub:1}

The system dynamics \eqref{eq:5}, after compensating in
feedforwarding for the known disturbances $D$, can be arbitrary
shaped by the state feedback $-Kx$, provided the control gains $K
\in \mathbb{R}^{4 \times 1}$ are designed appropriately. That
means the new system matrix $A^* = A-BK$ of the state feedback
closed-loop system must be Hurwitz, cf. e.g. \cite{antsaklis2007}.
Furthermore, $A^*$ should admit for the real eigenvalues only, in
order to compensate for undesired output oscillations. Note that
below, we will consider the state-feedback control part only, i.e.
without any pre-filter, correspondingly forward gain applied to
the reference value $r$. This is justified since our main focus
(in this section) is on stability and compensation of the
oscillations, and not on an accurate reference tracking.

Since $y(t)$ is the only available system measurement, a natural
way to keep usage of the state-feedback control is to design an
asymptotic state observer \cite{luenberger1971}, also well known
as Luenberger (or Luenberger type) observer. The system
\eqref{eq:5} proves to be fully observable, see e.g.
\cite{antsaklis2007,franklin2020}, so that an observation gain $Q
\in \mathbb{R}^{4 \times 1}$ can be determined so as to provide
estimate $\tilde{x}$ of the state vector.
\begin{figure}[!h]
\centering
\includegraphics[width=0.98\columnwidth]{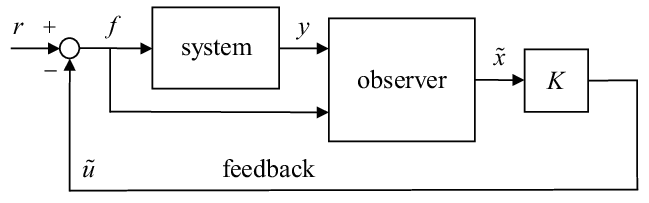}
\caption{Block diagram of the state-feedback with observer.}
\label{fig:4}
\end{figure}
The corresponding block diagram of a state-feedback control with
an observer is then shown in Fig. \ref{fig:4}. Recall that for an
asymptotically stable observation error $e(t) =
x(t)-\tilde{x}(t)$, i.e. for
$$
\underset{t\rightarrow\infty} {\lim} \| e(t)  \| = 0,
$$
the system matrix of the observation error dynamics
$$
\dot{e}(t) = \tilde{A} e(t) = (A-QC) e(t),
$$
must be Hurwitz. For ensuring the asymptotic observer operates
efficiently in combination with the state-feedback control, the
corresponding poles of $\tilde{A}$ are required (generally) to be
significantly faster than those of $A^*$. Also recall that once
the state-feedback, which includes now also the observer, is
closed, cf. Fig. \ref{fig:4}, the estimate dynamics becomes
\begin{equation}
\dot{\tilde{x}} = (A - BK -QC) \tilde{x} + Bf + Qy. \label{eq:41}
\end{equation}

Despite the well-known separation principle when designing the
Luenberger observer, that allows the poles of both observer and
state-feedback to be assigned independently of each other, a
practical realization of the closed-loop as in Fig. \ref{fig:4}
reveals less feasible for the system class introduced in section
\ref{sec:2:sub:1}. Next we will show it, although we should
emphasize that observer-based state feedback control as in Fig.
\ref{fig:4} is well established and accepted in many other types
of observable systems (also in applications).

\subsection{Practical infeasibility} \label{sec:3:sub:2}

Even though the designed observer and state-feedback loop based on
it have both a stable dynamics of the poles, we are to address
additional stability features, that from a loop transfer function
point of view, cf. Fig. \ref{fig:4}. The observer-based open-loop
transfer function from $r(s)$ to $\tilde{u}(s)$ is given by
\begin{equation}
L_o(s) =   K \bigl(sI - A + BK + QC\bigr)^{-1} \, \bigl[B \, Q \bigr] \, \left[%
\begin{array}{c}
  1 \\
  G(s) \\
\end{array}%
\right], \label{eq:42}
\end{equation} with the identity matrix $I$ of an appropriate dimension and
$$
G(s)=C(sI-A)^{-1}B.
$$
Note that without use of observer, i.e. if the full state $x(t)$
were measurable, the open-loop transfer function yields as
\begin{equation}
L_m(s) =   K \bigl(sI - A)^{-1}B.  \label{eq:43}
\end{equation}
Further we note that in the both above cases, the open-loop
transfer functions are not including the additional (disturbing)
actuator dynamics $F(s)$, cf. \eqref{eq:4}. Thus, also the
corresponding open-loop transfer functions $F(s)L_o(s)$ and
$F(s)L_m(s)$, respectively, must be inspected when analyzing the
practical infeasibility. Now, let us make use of the so-called
stability margin, or \emph{maximum sensitivity} see e.g.
\cite{aastrom2021} for details, which is defined as the maximum
magnitude, i.e.
\begin{equation}
S_{\max} = \underset{\Omega} {\max} \, \bigl| S(i\Omega) \bigr| =
\underset{\Omega} {\max} \, \Bigl| \bigl(1 + L(i\Omega)
\bigr)^{-1} \Bigr|, \label{eq:44}
\end{equation}
of the corresponding sensitivity function $S(\cdot)$. Recall that
the latter is directly related to the open-loop transfer function
$L(\cdot)$. Both are evaluated in frequency domain, where $\Omega$
in the angular frequency variable and $i$ is the imaginary unit
with $i^2 = -1$. Also recall that $S_{\max}$ indicates how close
the Nyquist plot of the open-loop transfer function bypasses from
the right the critical point $(-1, 0)$ in the complex plane, cf.
e.g. \cite{aastrom2021}. Thus, it represents, to say, a stability
capacity of the closed-loop system and is typically required to be
$S_{\max} < 2 \approx 6$ dB, cf. \cite{skogestad2005}. Systems
that have the loop transfer function with $S_{\max} > 4 \approx
12$ dB indicate a poor performance as well as poor robustness, cf.
\cite{skogestad2005}. Further we note that for a closed-loop with
the structure as in Fig. \ref{fig:4}, the sensitivity function
represents the transfer characteristics between the reference
value and the input to the system plant.

\begin{figure}[!h]
\centering
\includegraphics[width=0.98\columnwidth]{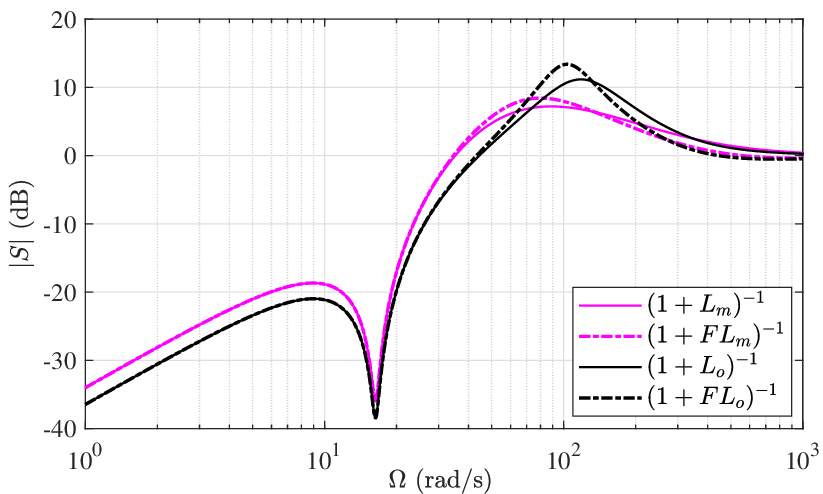}
\caption{Sensitivity function of the state-feedback control loop
with and without the use of observer.} \label{fig:42}
\end{figure}

Now consider the design of the above observer and the surrounding
state feedback control loop for the system \eqref{eq:5} by pole
placement. The exemplary assigned poles of the state feedback
control are $\lambda_c = \{-40, -42, -44, -60\} $ and those of the
asymptotic state observer are $\lambda_o = \{-498, -503, -508,
-513\}$. Note that all poles are real and placed sufficiently
close to each other, for the observer and controller respectively,
thus providing an approximately same time-scale of the natural
behavior of the all corresponding states and their estimates.
Here, the observer poles are approximately two and a half times
faster than those of the closed-loop. All four sensitivity
functions, i.e. with and without the use of observer and also with
and without the actuator dynamics, are shown in Fig. \ref{fig:42}.
One can recognize that already the state-feedback without observer
has a poor stability margin. When using observer, the $S_{\max}$
peak is further growing and becomes sharper. In case of the
actuator dynamics it reaches even 13.4 dB. Recall that a step
reference will excite all frequencies, so that the input
constraints (cf. section \ref{sec:2:sub:1}) can then be violated
during the transients.

\begin{figure}[!h]
\centering
\includegraphics[width=0.98\columnwidth]{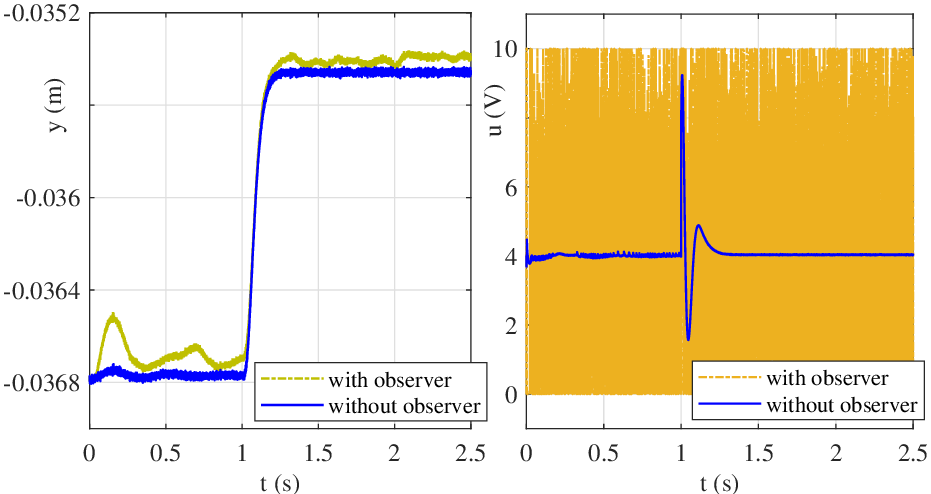}
\caption{Simulated response with and without the use of observer.}
\label{fig:43}
\end{figure}
The implemented model \eqref{eq:1}--\eqref{eq:5} is used in a
numerical simulation of the state-feedback control with and
without observer designed as above. The simulated output $y(t)$ is
subject to a minor (lower than in the experimental system)
measurement noise. Also the step reference $r(t)$ is chosen so
that the state-feedback control without observer is not saturated,
cf. section \ref{sec:2:sub:2}. The fixed-step solver with the
sampling time 0.0002 sec (same as in the real-time experiments) is
used. The output response and the control value of the simulation
are shown for both cases in Fig. \ref{fig:43}. One can see that in
case of observer, the control value comes permanently in a
high-frequent saturated behavior, that makes the observer-based
control practically infeasible.

%%%%%%%%%%%%%%%%%%%%%%%%%%%%%%%%%%%%%%%%%%%%%%%%%%%%%%%%%%%%%%%%%%%%%%%%%%%%%%%%
\section{Time delay based control} \label{sec:4}

\subsection{Time delay based control} \label{sec:4:sub:1}

The time delay based feedback control of the fourth-order
oscillatory systems, initially proposed in \cite{ruderman2021},
was introduced and analyzed in detail in \cite{ruderman23}, also
providing an experimental evaluation in combination with a
standard PI (proportional-integral) control. The time delay based
control
\begin{equation}\label{eq:4:01}
u_d(t) = \alpha \bigl(y(t)-y(t-\theta)\bigr),
\end{equation}
relies on the knowledge of the oscillation frequency $\omega$, and
assumes the time delay constant
\begin{equation}
\theta = - \arg \bigl[G(i \omega)\bigr] \omega^{-1},
\label{eq:4:02}
\end{equation}
and the gaining factor $\alpha > 0$. The latter is the design
parameter. The system transfer function is given by
\begin{equation}
G(i \Omega) = \frac{y(i \Omega)}{u(i \Omega)} = C \bigl( i \Omega
I - A \bigr)^{-1} B. \label{eq:4:02}
\end{equation}

\begin{figure}[!h]
\centering
\includegraphics[width=0.98\columnwidth]{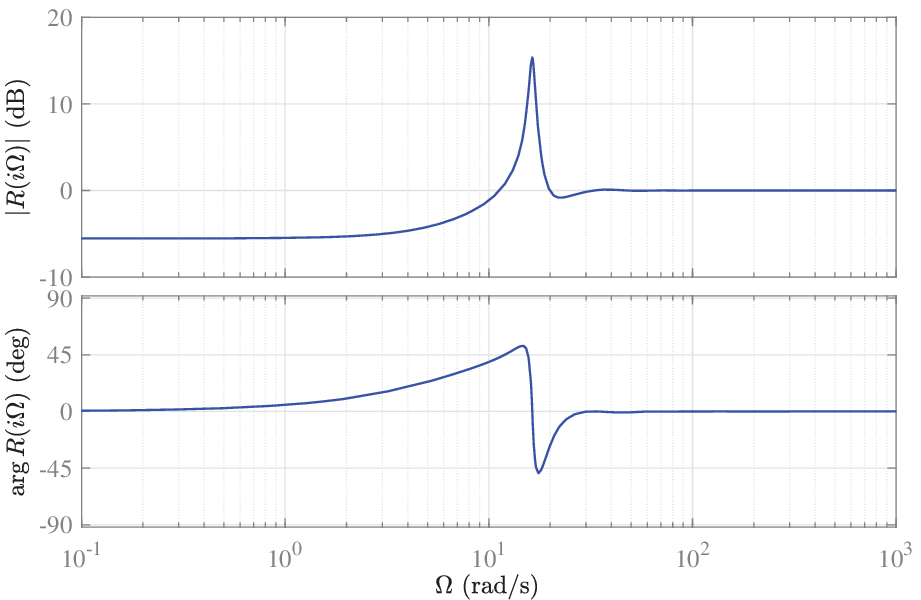}
\caption{Reshaping \eqref{eq:4:04} of the system transfer
characteristics.} \label{fig:401}
\end{figure}

We note that the time delay based control \eqref{eq:4:01} is
largely attenuating the system resonance peak around $\omega_0$,
without much reshaping the $G(i \Omega)$ transfer characteristics
at other frequencies. Expressing the transfer function
\eqref{eq:4:02} as a ratio $G(i \Omega)=N(i \Omega) P(i
\Omega)^{-1}$ of the corresponding polynomials $N(\cdot)$ and
$P(\cdot)$, and rewriting \eqref{eq:4:01} in frequency domain as
\begin{equation}
U_d(i \Omega) = \alpha \Bigl (1 - \exp \bigl(- i \Omega \: \theta
\bigr) \Bigr), \label{eq:4:03}
\end{equation}
one can show that the closed-loop $G_{cl} = N (P-N U_d)^{-1}$ is
reshaping the system plant transfer characteristics as
\begin{equation}
R (i \Omega) = \frac{G(i \Omega)}{G_{cl}(i \Omega)} = 1 -
\frac{N(i \Omega) U_d (i \Omega)}{P(i \Omega)}.\label{eq:4:04}
\end{equation}
The reshaping characteristics \eqref{eq:4:04} of the system
frequency response, here without the constant disturbance $D$ and
assuming $\tau=0$ in \eqref{eq:4}, is exemplary shown in Fig.
\ref{fig:401} for $\alpha = 100$.
One can recognize that the principal difference between the
original system plant $G(i \Omega)$ and that one with the time
delay based compensator, i.e. $G_{cl}(i \Omega)$, is precisely the
resonance peak of $G(i \Omega)$. At higher frequencies, there is
no changes in the amplitude response, while at lower frequencies
an acceptable gain reduction is about $-5$ dB. This can then be
taken into account when the compensated system $G_{cl}(i \Omega)$
will be closed by an outer tracking control, cf.
\cite{ruderman23}. One can also recognize that the phase response
of $G(i \Omega)$ and $G_{cl}(i \Omega)$ are essentially the same
to the left- and right-hand-side of the resonance frequency.

\subsection{Robust frequency estimator} \label{sec:4:sub:2}

The robust frequency estimator, proposed in \cite{ruderman2022},
can be used for an online tuning of the $\omega$-parameter,
provided the measured oscillatory signal $w(t)$ in unbiased. The
estimator dynamics is given by, cf. \cite{ruderman2022},
\begin{eqnarray}
\nonumber \left(%
\begin{array}{c}
  \dot{\eta}_1 \\
  \dot{\eta}_2 \\
\end{array}%
\right) & = & \left(%
\begin{array}{cc}
  0 & 1 \\
  -\tilde{\omega}^2 & -2 \tilde{\omega} \\
\end{array}%
\right) \left(%
\begin{array}{c}
  \eta_1 \\
  \eta_2 \\
\end{array}%
\right) + \left(%
\begin{array}{c}
  0 \\
  2 \tilde{\omega} \\
\end{array}%
\right) w,
\\[1mm]
\nu & = & \left(%
\begin{array}{cc}
  0 & 1 \\
\end{array}%
\right) \left(%
\begin{array}{c}
  \eta_1 \\
  \eta_2 \\
\end{array}%
\right), \label{eq:4:1}
\end{eqnarray}
with the frequency-estimate adaptation law given by
\begin{equation}
\dot{\tilde{\omega}} = -\gamma \, \tilde{\omega} \,
\mathrm{sign}(\eta_1)(w - \nu). \label{eq:4:2}
\end{equation}
Here $\gamma > 0$ is the gain parameter of the estimation. Note
that the right-hand-side of \eqref{eq:4:2} includes (additionally
in comparison to \cite{ruderman2022}) a multiplication with
$\tilde{\omega}$, so as to avoid the frequency estimate
$\tilde{\omega}$ bypassing into the negative range. For details on
the stability and performance of the robust frequency estimator
the reader is referred to \cite{ruderman2022}.

In order for a biased oscillation output $y(t)$ can equally be
used in the frequency estimator \eqref{eq:4:1}, \eqref{eq:4:2},
the following dynamic bias-canceling is proposed
\begin{equation}
w(t) = y(t) - y\Bigl(t-\frac{\pi}{\beta}\Bigr), \quad \omega <
\beta < 3 \omega, \label{eq:4:3}
\end{equation}
where $\beta$ is a free adjustable time-delay parameter. Assuming
a biased, by the term $Y_0$, harmonic oscillation
\begin{equation}
y(t) = Y_0 + Y \sin \bigl(\omega t  + \phi \bigr), \label{eq:4:4}
\end{equation}
and substituting it into \eqref{eq:4:3} results in
\begin{equation}
w(t) = Y \biggl( \sin \bigl(\omega t  + \phi \bigr) - \sin
\Bigl(\omega t + \phi - \omega \frac{\pi}{\beta} \Bigr) \biggr).
\label{eq:4:5}
\end{equation}
One can recognize that \eqref{eq:4:5} constitutes also a harmonic
signal with the same fundamental frequency $\omega$. The signal is
unbiased and has another amplitude and phase comparing to the
harmonic part of \eqref{eq:4:4}. Also worth noting is that $w(t)$
is not zero signal as long as $\omega \pi \beta^{-1} \neq 0$. If
some nominal or upper-bound value of $\omega$ is known, the phase
shifting factor $\beta$ can be assigned in a relatively large
range, cf. \eqref{eq:4:3}.
\begin{figure}[!h]
\centering
\includegraphics[width=0.98\columnwidth]{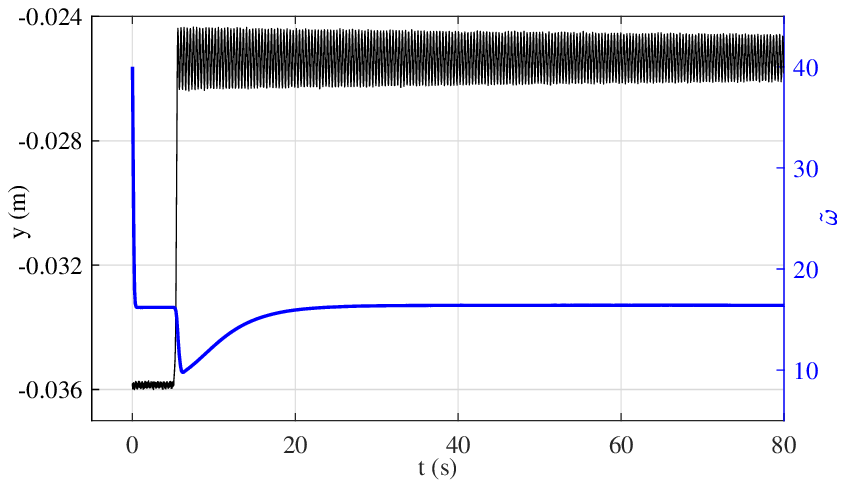}
\caption{Converging $\tilde{\omega}(t)$ versus the measured
oscillating $y(t)$.} \label{fig:41}
\end{figure}
Since the robust estimator \eqref{eq:4:1}, \eqref{eq:4:2} is
unsensitive to both, the phase $\phi$ and slow variations of $Y$,
see \cite{ruderman2022}, the bias-freed input \eqref{eq:4:5} can
directly be used for estimation of $\tilde{\omega}$. Recall that
$\tilde{\omega}(t)$ has a convergence behavior which rate is
controllable by $\gamma$. An exemplary convergence of
$\tilde{\omega}$ is shown in Fig. \ref{fig:41}, together with the
used $y(t)$ measurement. Note that the measured output is biased
before and after the step-wise excitation. The initial value is
set to $\tilde{\omega}(0)=40$ and the gain to $\gamma=200$. One
can recognize that after the exciting transient of $y(t)$, the
$\tilde{\omega}(t)$ is re-converging towards its final value
$\approx 16.4$ rad/sec.

%%%%%%%%%%%%%%%%%%%%%%%%%%%%%%%%%%%%%%%%%%%%%%%%%%%%%%%%%%%%%%%%%%%%%%%%%%%%%%%%
\section{Experimental control evaluation} \label{sec:5}

A standard PI feedback controller
\begin{equation}
u_{pi}(t) = K_p e(t) + K_i \int e(t) dt, \label{eq:5:1}
\end{equation}
operating on the output error $e(t) = r(t)-y(t)$, with $r(t)$ to
be the set reference value for the oscillatory load, can be
desirable for the following reasons. (i) The single available
system state is $y(t)$. (ii) An integral control action is
required for guaranteeing a steady-state accuracy. (iii) An
additional differential control action (i.e. resulting in PID
control) is not contributing to stabilization of the oscillatory
load in the fourth-order system \eqref{eq:5}. (iv) A
state-feedback control, that requires an additional state
observer, fails practically for the systems \eqref{eq:2},
\eqref{eq:3}, as discussed in detail in section \ref{sec:3}. At
the same time, one can show that the open-loop transfer function
$L(i\Omega) =y(s)/r(s) = PI(i\Omega)G(i\Omega)$, where
$PI(i\Omega)$ is the transfer function corresponding to
\eqref{eq:5:1}, has a marginal or even none gain margin. For the
basics on gain margin and loop transfer function analysis we refer
to e.g. \cite{franklin2020}. Also the so-called maximum
sensitivity, cf. e.g. \cite{aastrom2021}, will have a relatively
high number for the corresponding $S(i\Omega) =
\bigl(1+L(i\Omega)\bigr)^{-1}$, cf. with the analysis made in
section \ref{sec:3}. This indicates a low stability of the
closed-loop system.

The overall control law evaluated experimentally is
\begin{equation}
u(t) = u_{pi}(t) + u_d(t) + 4.035, \label{eq:5:2}
\end{equation}
where the last constant term of the right-hand-side compensates
for the known gravity disturbance, cf. \eqref{eq:2}, \eqref{eq:3},
\eqref{eq:4}. Note that for the assigned $K_p=100$, $K_i = 170$,
cf. \cite{ruderman23}, the loop transfer function $L(\cdot)$
without \eqref{eq:4:03} has a sufficient phase margin of 46 deg,
but the missing gain margin of $-4.2$ dB.

\begin{figure}[!h]
\centering
\includegraphics[width=0.98\columnwidth]{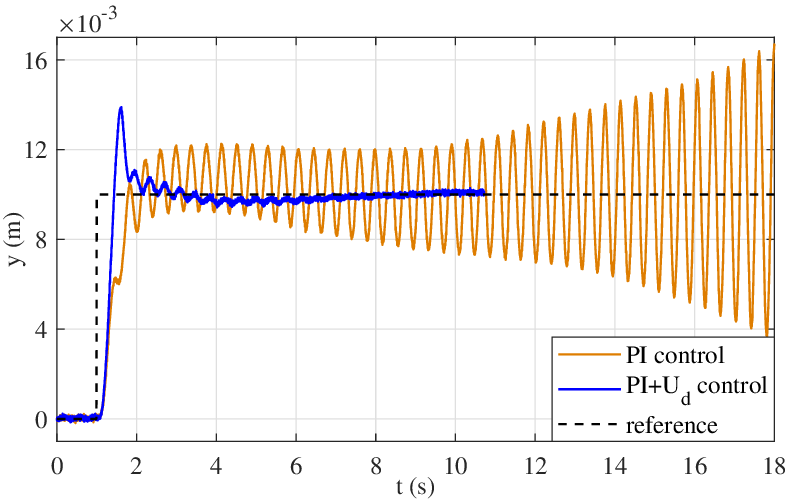}
\caption{Measured $y(t)$ controlled by $PI$ (i.e. $\alpha = 0$)
and $PI+U_d$ controllers, with $\alpha = 100$ and the fixed
$\theta$ value.} \label{fig:51}
\end{figure}

First, the closed-loop response controlled with \eqref{eq:5:2} is
experimentally evaluated as shown in Fig. \ref{fig:51}, once
without the time delay control part (i.e. with $\alpha = 0$) and
once with the time delay control part with $\alpha = 100$. Note
that here a fixed time delay constant, cf. section
\ref{sec:4:sub:1}, is assigned from the known system parameter
$\omega$.
While the time delay based control provides a relatively fast
cancelation of the otherwise oscillating output, cf. also with
Fig. \ref{fig:3}, the pure PI control drives the system to a
visible instability over the time.

\begin{figure}[!h]
\centering
\includegraphics[width=0.98\columnwidth]{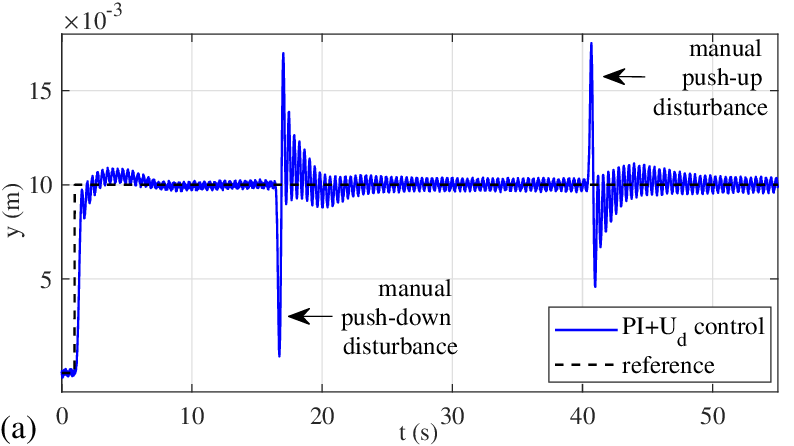}
\includegraphics[width=0.98\columnwidth]{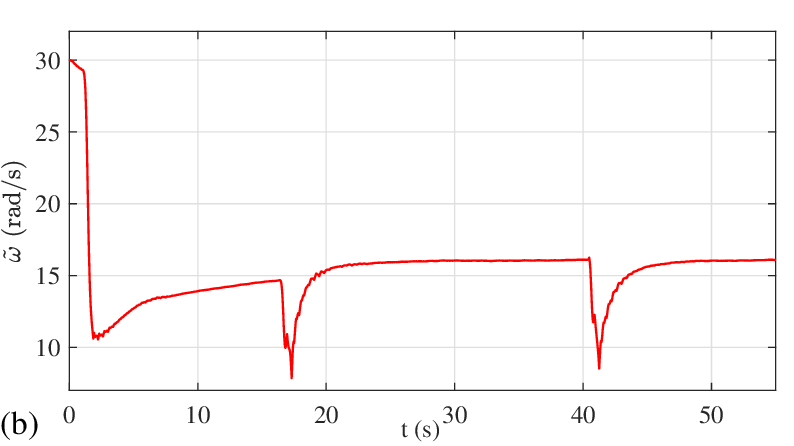}
\caption{Measured $y(t)$ controlled with $PI+U_d$ for $\alpha =
100$ and online adapted $\theta$ in (a), disturbances are marked;
the convergence of $\tilde{\omega}(t)$ in (b).}\label{fig:52}
\end{figure}

Next, the feedback control \eqref{eq:5:2} is experimentally
evaluated when applying an online adaptation of $\theta$ by means
of the robust frequency estimator described in section
\ref{sec:4:sub:2}. The adaptation gain is assigned to $\gamma =
600$. The results are shown in Fig. \ref{fig:52}, where the
controlled output in depicted in (a), and the time progress of the
$\tilde{\omega}(t)$ estimate is depicted in (b). Also the manual
mechanical disturbances, which are additionally exciting the
output oscillations, were applied, once by pushing down and once
by pushing up the passive load, see in Fig. \ref{fig:52} (a)
marked by the arrows.
One can recognize both, a stable attenuation of the output
oscillations and a robust convergence of the $\tilde{\omega}(t)$
estimate. Important to notice is also that due to some (not
modeled) nonlinear by-effects in the stiffness, the oscillation
frequency $\omega$ experiences certain fluctuations depending on
the amplitude of $y(t)$, that means on the elongation of the
connecting spring.

\begin{figure}[!h]
\centering
\includegraphics[width=0.98\columnwidth]{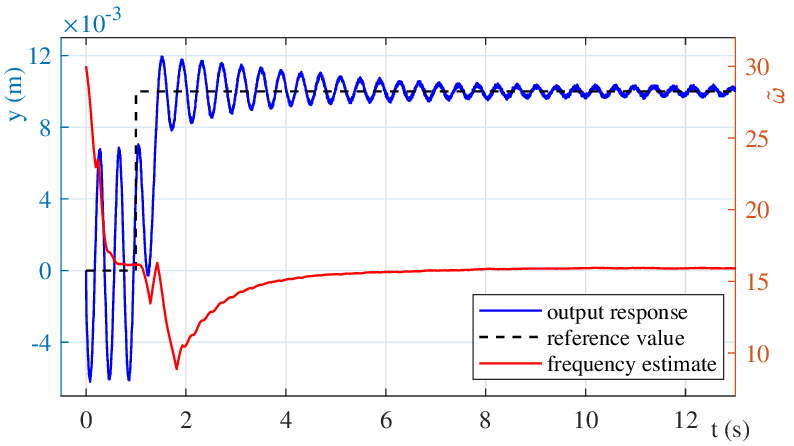}
\caption{Measured $y(t)$ controlled with $PI+U_d$ for $\alpha =
100$ and the oscillatory initial conditions, and the convergence
of the $\tilde{\omega}(t)$ estimate.} \label{fig:53}
\end{figure}

For further evaluating robustness of the adaptive control
\eqref{eq:5:2}, i.e. including online estimation of
$\tilde{\omega}$, the experiments are performed for oscillatory
initial conditions, see Fig. \ref{fig:53}. The control parameters
are the same as above. One can recognize a largely oscillating
$y(t)$ before the step reference is applied.
For largely (i.e. more pronounced) oscillations of $y(t)$, the
$\tilde{\omega}(t)$ estimate converges faster, which is then
perturbed by the output transient with $Y_0 \neq \mathrm{const}$,
cf. \eqref{eq:4:4}. After transient, the convergence of
$\tilde{\omega}(t)$ recovers.

%%%%%%%%%%%%%%%%%%%%%%%%%%%%%%%%%%%%%%%%%%%%%%%%%%%%%%%%%%%%%%%%%%%%%%%%%%%%%%%%
\section{Conclusions} \label{sec:6}

In this paper, we have addressed the fourth-order non-collocated
systems with low-damped oscillating passive loads. Approaching the
real applications, an actuator body is subject to the input and
output constraints. We analyzed and demonstrated numerically that
an observer-based state-feedback control reveals practically
infeasible, despite the system dynamics proves to be observable.
As a robust alternative, the time delay based control
\cite{ruderman2021,ruderman23} was applied for stabilizing the
otherwise unstable PI controller of the set reference. The
bias-canceling extension of the robust frequency estimator
\cite{ruderman2022} was introduced, which allows for an online
(adaptive) tuning of the time delay based controller. Various
dedicated experiments were shown as confirmatory.

%%%%%%%%%%%%%%%%%%%%%%%%%%%%%%%%%%%%%%%%%%%%%%%%%%%%%%%%%%%%%%%%%%%%%%%%%%%%%%%%
\bibliographystyle{IEEEtran}
\bibliography{references}

\begin{thebibliography}{10}
\providecommand{\url}[1]{#1}
\csname url@rmstyle\endcsname
\providecommand{\newblock}{\relax}
\providecommand{\bibinfo}[2]{#2}
\providecommand\BIBentrySTDinterwordspacing{\spaceskip=0pt\relax}
\providecommand\BIBentryALTinterwordstretchfactor{4}
\providecommand\BIBentryALTinterwordspacing{\spaceskip=\fontdimen2\font plus
\BIBentryALTinterwordstretchfactor\fontdimen3\font minus
  \fontdimen4\font\relax}
\providecommand\BIBforeignlanguage[2]{{%
\expandafter\ifx\csname l@#1\endcsname\relax
\typeout{** WARNING: IEEEtran.bst: No hyphenation pattern has been}%
\typeout{** loaded for the language `#1'. Using the pattern for}%
\typeout{** the default language instead.}%
\else
\language=\csname l@#1\endcsname
\fi
#2}}

\bibitem{vaughan2010}
J.~Vaughan, D.~Kim, and W.~Singhose, ``Control of tower cranes with
  double-pendulum payload dynamics,'' \emph{IEEE Transactions on Control
  Systems Technology}, vol.~18, no.~6, pp. 1345--1358, 2010.

\bibitem{wang2023}
J.~Wang and W.~T. van Horssen, ``On resonances and transverse and longitudinal
  oscillations in a hoisting system due to boundary excitations,''
  \emph{Nonlinear Dynamics}, vol. 111, pp. 5079--5106, 2023.

\bibitem{besselink2015}
B.~Besselink, T.~Vromen, N.~Kremers, and N.~Van De~Wouw, ``Analysis and control
  of stick-slip oscillations in drilling systems,'' \emph{IEEE Trans. on
  Control Systems Technology}, vol.~24, no.~5, pp. 1582--1593, 2015.

\bibitem{ruderman2021}
M.~Ruderman, ``Robust output feedback control of non-collocated low-damped
  oscillating load,'' in \emph{IEEE 29th Mediterranean Conference on Control
  and Automation (MED)}, 2021, pp. 639--644.

\bibitem{ruderman23}
M.~Ruderman, ``Time-delay based output feedback control of fourth-order
  oscillatory systems,'' \emph{Mechatronics}, vol.~94, p. 103015, 2023.

\bibitem{kao2004}
C.-Y. Kao and B.~Lincoln, ``Simple stability criteria for systems with
  time-varying delays,'' \emph{Automatica}, vol.~40, no.~8, pp. 1429--1434,
  2004.

\bibitem{fridman2006input}
E.~Fridman and U.~Shaked, ``Input--output approach to stability and l2-gain
  analysis of systems with time-varying delays,'' \emph{Systems \& Control
  Letters}, vol.~55, no.~12, pp. 1041--1053, 2006.

\bibitem{gu2003}
K.~Gu, V.~Kharitonov, and J.~Chen, \emph{Stability of time-delay
  systems}.\hskip 1em plus 0.5em minus 0.4em\relax Springer, 2003.

\bibitem{michiels2014}
W.~Michiels and S.-I. Niculescu, \emph{Stability, control, and computation for
  time-delay systems: an eigenvalue-based approach}.\hskip 1em plus 0.5em minus
  0.4em\relax SIAM, 2014.

\bibitem{fridman2014}
E.~Fridman, ``Tutorial on lyapunov-based methods for time-delay systems,''
  \emph{Eur. Journal of Control}, vol.~20, no.~6, pp. 271--283, 2014.

\bibitem{liberzon2003}
D.~Liberzon, \emph{Switching in systems and control}.\hskip 1em plus 0.5em
  minus 0.4em\relax Springer, 2003.

\bibitem{ruderman2022}
M.~Ruderman, ``One-parameter robust global frequency estimator for slowly
  varying amplitude and noisy oscillations,'' \emph{Mechanical Systems and
  Signal Processing}, vol. 170, p. 108756, 2022.

\bibitem{voss2022}
B.~Vo{\ss}, M.~Ruderman, C.~Weise, and J.~Reger, ``Comparison of
  fractional-order and integer-order {H}-infinity control of a non-collocated
  two-mass oscillator,'' \emph{IFAC-PapersOnLine}, vol.~55, pp. 145--150, 2022.

\bibitem{antsaklis2007}
P.~Antsaklis and A.~Michel, \emph{A Linear Systems Primer}.\hskip 1em plus
  0.5em minus 0.4em\relax Birkh{\"a}user Boston, 2007.

\bibitem{luenberger1971}
D.~Luenberger, ``An introduction to observers,'' \emph{IEEE Transactions on
  Automatic Control}, vol.~16, no.~6, pp. 596--602, 1971.

\bibitem{franklin2020}
G.~Franklin, J.~Powell, and A.~Emami-Naeini, \emph{Feedback control of dynamic
  systems}.\hskip 1em plus 0.5em minus 0.4em\relax Pearson, 2020.

\bibitem{aastrom2021}
K.~J. {\AA}str{\"o}m and R.~M. Murray, \emph{Feedback systems: an introduction
  for scientists and engineers}.\hskip 1em plus 0.5em minus 0.4em\relax
  Princeton University Press, 2021.

\bibitem{skogestad2005}
S.~Skogestad and I.~Postlethwaite, \emph{Multivariable feedback control:
  analysis and design}.\hskip 1em plus 0.5em minus 0.4em\relax John Wiley \&
  Sons, 2005.

\end{thebibliography}

\end{document}